\documentclass[a4paper]{article}

\usepackage[english]{babel}
\usepackage[utf8x]{inputenc}
\usepackage{amsmath}
\usepackage{graphicx}
\usepackage[colorinlistoftodos]{todonotes}
\usepackage{amsthm}
\usepackage{bigints} 

\usepackage[a4paper, total={6.5in, 10in}]{geometry}

\usepackage[english]{babel}
\usepackage[utf8x]{inputenc}
\usepackage{amsfonts}
\usepackage{graphicx}
\usepackage{float}
\usepackage{hyperref}
\usepackage{url}

\newcommand{\dzbar}{\partial_{\bar z}}

\usepackage{geometry}
\providecommand{\keywords}[1]{\textbf{\textit{Key words:}} #1}

\begin{document}
\title{
\begin{flushright}\ \vskip -2cm {\small {DAMTP-2016-83}}\end{flushright}
\vskip 15pt
\bf{Integrable Abelian Vortex-like Solitons}
\vskip 30pt}
\author{
Felipe Contatto$^{1,2}$\\[30pt]
{\em \normalsize
$\phantom{}^1$ Department of Applied Mathematics and Theoretical Physics,}\\[0pt] 
{\em \normalsize University of Cambridge, Wilberforce Road, 
Cambridge CB3 0WA, U.K. }\\[10pt] 
{\em \normalsize
$\phantom{}^2$ CAPES Foundation, Ministry of Education of Brazil, }\\[0pt] 
{\em \normalsize Bras\'ilia - DF 70040-020, Brazil.}\\[10pt] 
{\small Email: felipe.contatto@damtp.cam.ac.uk}\\[5pt]
}
\date{February 2017}
\maketitle
\vskip 20pt

\begin{abstract}
We propose a modified version of the Ginzburg-Landau energy functional admitting static solitons and determine all the Painlevé-integrable cases of its Bogomolny equations of a given class of models. Explicit solutions are determined in terms of the third Painlevé transcendents, allowing us to calculate physical quantities such as the vortex number and the vortex strength. These solutions can be interpreted as the usual Abelian-Higgs vortices on surfaces of non-constant curvature with conical singularity.
\end{abstract}

\section{Introduction}

The Ginzburg-Landau model is a relativistic $U(1)$-gauge theory with a $\phi^4$ potential for the Higgs field, which is relevant in the theory of superconductors near the critical coupling. This model admits topological solitons called vortices stabilised by the topological charge. The existence and analyticity properties of Abelian vortices on the plane were largely studied in particular in \cite{JafTaubes,Taubes1980}. Generalised vortices were proposed by Lohe \cite{Lohe1981}, whose model admits other types of potentials at the expense of minimal coupling between the Higgs and gauge fields. In fact, Lohe's model modify the kinetic term of the Higgs field as well as the potential in such a way that the Bogomolny argument still holds. The existence of generalised vortices under an analytical point of view was established in \cite{LoheVDH1983}. 

Integrability of the Abelian-Higgs model is well known on a hyperbolic background where the general solution can be explicitly described in terms of holomorphic maps \cite{Witten1977,Strachan1992,ManRink2010,MalMan2015}. Moreover, Painlev\'e analysis shows these are the only cases in which vortices are integrable in the Painlev\'e sense \cite{Shiff1991}. However two more isolated integrable cases of Abelian vortices were found by allowing the background metric to depend on the Higgs field, allowing the Bogomolny equations to be written as sinh-Gordon and Tzitzeica equations \cite{MD2012}, albeit they do not arise from a variational principle approach from the Ginzburg-Landau model. 

In this paper we present another modified version of the Abelian-Higgs model by coupling the Yang-Mills term of the Lagrangian with the Higgs field through a continuous function denoted by $G(|\phi|)$. Under mild conditions on $G$ and upon a suitable modification of the potential energy, the model admits vortex-like topological solitons from the Bogomolny argument. A choice of coupling function of the form $G(|\phi|)=|\phi|^{q+1}$, where $q\in\mathbb R$, will give rise to a model that includes the usual Abelian-Higgs vortices as a particular case ($q=-1$), but admits further Painlev\'e-integrable cases, including those described in \cite{MD2012}, providing a variational approach to them. We determine all the possible values of $q$ and all possible background metrics yielding integrable models using the Painlev\'e test. 

In section \ref{secModifiedVortices} we present the modified Ginzburg-Landau Lagrangian and the corresponding Bogomolny equations, which can actually be reduced to a single PDE that will be referred to as the modified Taubes equations. On surfaces of revolution, this equation admits a symmetry reduction to an ODE by rotational symmetry around the origin. On a surface of revolution, if the PDE passes the Painlev\'e test then so does the reduced ODE, but nothing guarantees that the converse is true.
Thus we perform a separated analysis at the reduced ODE in section \ref{secODE} and at the PDE in section \ref{secPDE}. It turns out that the analysis of the ODE does not reveal more integrable cases than that of the PDE.

Within the class of models considered, there can only be integrable cases on a hyperbolic background with Gauss curvature $-1/2$ and on a flat surface. On a hyperbolic surface, the only integrable case corresponds to $q=-1$, which is the usual Abelian-Higgs model. The converse was already known \cite{Shiff1991}. On a flat background, there are three integrable models, corresponding to $q=1/3$, $q=0$ and $q=-1/3$. The first two values are equivalent to the isolated integrable vortices of \cite{MD2012} while $q=-1/3$ gives rise to a new solution and completes the list of integrable models under the class we consider. The Bogomolny equation for this case can be written as the Tzitzeica equation. In order to write explicit soliton solutions with finite energy, some boundary conditions apply. It is not obvious how to apply these boundary conditions to solutions of the Tzitzeica equation, but if we restrict to rotationally symmetric solutions and reduce the Tzitzeica PDE to a Painlev\'e III ODE whose solutions are well known in the asymptotics \cite{Kit1989}, we can impose the right boundary conditions by fixing some parameters of the third Painlev\'e transcendents. Besides its integrability features, these solutions can be interpreted \cite{MD2012} as usual Abelian-Higgs vortices on backgrounds with conical and curvature singularities at the origin. It is worth noticing that even though our model includes the usual Abelian-Higgs model, there are other types of similar integrable vortex equations on different backgrounds \cite{Popov2012,Manton2013,Manton2016} that it does not cover.

\section{Modified Abelian-Higgs model}\label{secModifiedVortices}

We start with the Ginzburg-Landau theory with a modified Lagrangian on a smooth manifold $\mathbb R\times \Sigma$ with Lorentzian  metric $ds^2=dt^2-\Omega(dx^2+dy^2)$,
\begin{equation}\label{modLag}
L=\int\left(-\frac{G(|\phi|)^2}{4}F_{\mu\nu}F^{\mu\nu}+\frac{1}{2}\overline{D_\mu\phi} D^\mu\phi-V(|\phi|)\right)\Omega\, d^2x,
\end{equation}
where $G$ is a continuous function of $|\phi|$ on its domain of definition, $D_\mu=\partial_\mu-i a_\mu$ is the covariant derivative and $F_{\mu\nu}=\partial_\mu a_\nu-\partial_\nu a_\mu$ is the curvature $2$-form of the $U(1)$-connection $a$. The space indices will be denoted by $i,j,k\dots$ and range from $1$ to $2$ as $(x^1,x^2)=(x,y)$. We will also use complex coordinates $z=x+iy$ and polar coordinates $z=r e^{i\theta}$, whenever it is convenient.

If the potential is $V(|\phi|)=\frac{1}{8 G(|\phi|)^2}\left(1-|\phi|^2\right)^2$, which differs from the Ginzburg-Landau $\phi^4$ theory by the factor $G(|\phi|)^2$ in the denominator and spontaneously breaks symmetry, then the usual Bogomolny argument can be applied.  In fact, the modified energy functional is
\begin{align}\label{modEn}
E&=\frac{1}{2}\int\left(\frac{G(|\phi|)^2}{\Omega^2}B^2+\overline{D_i\phi} D^i\phi+\frac{1}{4G(|\phi|)^2}\left(1-|\phi|^2\right)^2\right)\Omega d^2x \nonumber\\
&=\frac{1}{2}\int\left[\frac{G(|\phi|)^2}{\Omega}\left(B-\frac{\Omega}{2G(|\phi|)^2}(1-|\phi|^2)\right)^2+|D_{\bar z}\phi|^2+B-i\left(\partial_1(\bar \phi D_2\phi)-\partial_2(\bar \phi D_1\phi)\right)\right] d^2x\nonumber\\
&=\frac{1}{2}\int\left[ \frac{G(|\phi|)^2}{\Omega}\left(B-\frac{\Omega}{2G(|\phi|)^2}(1-|\phi|^2)\right)^2+|D_{\bar z}\phi|^2\right] d^2x+\pi N,
\end{align}
where $B$ denotes the component $F_{12}$ and $N\equiv\frac{1}{2\pi}\int_\Sigma B$ is supposed to be positive, as the case $N<0$ is analogous. We have used the boundary conditions $|\phi|\to 1$ and $D_i\phi\to 0$ as $z$ approaches the boundary of $\Sigma$ in the last equality. We assume that all terms in the energy functional are integrable so that it is well defined, which is true for the  integrable cases analysed here.

Thus the modified Bogomolny equations are
\begin{align}
&D_{\bar z}\phi\equiv\dzbar\phi-ia_{\bar z}\phi=0\label{modBog1} \\
&B=\frac{\Omega}{2G(|\phi|)^2}(1-|\phi|^2).\label{modBog2}
\end{align}
Eliminating $a_{\bar z}$ in the second equation using the first one, remembering that $a_{\bar z}=\overline{a_z}$, gives the modified Taubes equation,
\begin{equation}\label{modTaubes}
\Delta_0 h+\frac{\Omega}{G(e^{h/2})^2}\left(1-e^h\right)=0,
\end{equation}
where $h=\ln |\phi|^2$ and $\Delta_0=\partial_x^2+\partial_y^2$ is the Laplacian operator. Solving (\ref{modTaubes}) and imposing the boundary conditions above gives rise to vortex-like topological solitons the a surface $\Sigma$ defined by constant time slices with metric
$$
g=\Omega (dr^2+r^2d\theta^2),
$$
and Gauss curvature given by
$$
K_\Sigma=-\frac{1}{2\Omega}\Delta_0\ln\Omega.
$$

Notice that equation (\ref{modTaubes}) should be modified in case $\phi$ has zeros, as this implies the presence of logarithmic singularities for $h$ and the term $\Delta_0 h$ would generate delta functions. In fact, the usual Taubes equation ($G=1$) is often corrected with delta function sources added by hand to take these singularities into account \cite{ManSutbook}, which may occur at a general point and are the coordinates of the moduli space of vortices. This means that these log-singularities are movable in general and we will bypass them in our Painlev\'e analysis using an exponential change of variables $\chi=e^h$ in the following sections.

From now on, as explained in the Introduction, we will assume that $G(e^{h/2})^2=e^{(q+1)h/2}$, for a general $q\in\mathbb R$ and, in the next sections, study the integrability features of equation (\ref{modTaubes}). With this choice, we are going to impose another two conditions to the Higgs field. First, we require that the Higgs field is non-vanishing except on a finite number $l$ of distinct points $z_1,\dots,z_l$ and secondly that in a neighbourhood of each point $z_i$, there exists $n_i\in\mathbb N^*$ such that 
\begin{equation}\label{eqBCond}
\phi=(z-z_i)^{n_i}\psi_i(z,\bar z),
\end{equation}
where $\psi_i$ is a continuous function on the neighbourhood that is differentiable everywhere except possibly at $z_i$.

These conditions are the most natural ones to impose when seeking a generalisation of the Abelian Higgs model. In fact, they are immediately satisfied for the Abelian Higgs model, which can be proved from the existence of smooth solutions to the Bogomolny equations \cite{JafTaubes}; however, smoothness is an excessively strong condition to impose on the solutions of (\ref{modBog1}-\ref{modBog2}) in general. We will justify these conditions in section \ref{secODE} in order to rule out solutions on smooth surfaces that do not have a similar behaviour as Abelian vortices (cf. (\ref{solLiou1}-\ref{solLiou3})).

To begin with, we will suppose that $\Sigma$ is a surface of revolution so that the conformal factor is only a function of the radial coordinate, $\Omega=\Omega(r)$, as well as the modulus of the Higgs field, i.e., $h=h(r)$. This reduces (\ref{modTaubes}) to an ODE, that will be analysed in section \ref{secODE}. Then, in section \ref{secPDE}, we perform the analysis to the PDE (\ref{modTaubes}) in general.

\section{Painlev\'e analysis of the ODE}\label{secODE}
We apply Painlev\'e analysis \cite{ARS1980} to seek choices of $\Omega$ such that equation (\ref{modTaubes}) for $G(e^{h/2})^2=e^{(q+1)h/2}$ is integrable, assuming cylindrical symmetry, that is to say $\Omega=\Omega(r)$ and $h=h(r)$. Because of the logarithmic divergence of $h$ where the Higgs field vanishes, we look instead at the equation for $\chi=e^{h}$, for which the ODE reduced from (\ref{modTaubes}) is
\begin{equation}\label{modTaubeschi}
\chi''-\frac{\chi'^2}{\chi}+\frac{1}{r}\chi'+\frac{\Omega(r)}{\chi^{(q-1)/2}}\left(1-\chi\right)=0.
\end{equation}

In practise, the aim of the analysis is to determine in which cases the general solutions of the ODE can be locally written in the form $\chi=(r-r_0)^p\sum_{j=0}^\infty \chi_j (r-r_0)^j$, where $\chi_j$ are constants, $\chi_0\not=0$, $r_0 > 0$ is arbitrary and $p$ is assumed to be an integer, a hypothesis that will be justified in the next section. The arbitrary constant $r_0$ represents the position of a movable singularity (either of $\chi$ or $1/\chi$), that we expect not to be critical (or multivalued) for the Painlev\'e property to hold. We suppose that $r_0\not=0$ in order to avoid the coordinate singularity at $r=0$ of (\ref{modTaubeschi}). We look for the dominant behaviour by substituting $\chi\sim \chi_0\left(r-r_0\right)^p$ in (\ref{modTaubeschi}), we have to study it differently according to $p$ is positive or negative. 

We start by supposing that $p>0$. Balancing of the dominant terms (1st, 2nd and 4th) requires $p=\frac{4}{1+q} > 0$ and $\chi_0^{2/p}=\frac{\Omega(r_0)}{p}$. Since we are dealing with a second order ODE, its general solution should involve two constants of integration. One of them is the arbitrary constant $r_0$ itself, the other one will be a $\chi_{s}$, for some $s\geq 0$. The order $p+s$ in which this second constant appears is called the order of resonance (or the Fuchs index). Upon substituting the above series in (\ref{modTaubeschi}) and balancing all the powers of $r-r_0$, the constants $\chi_j$ should in principle be determined in terms of $r_0$ and $\chi_{s}$. Notice that the constant $\chi_0$ was fixed above in terms of $r_0$, so we can already tell that $s>0$. Moreover, $\chi_{s}$ will be a free parameter if and only if the leading order in which it appears in the expansion of (\ref{modTaubeschi}) involves $\chi_j$ algebraically for some $j>s$, so that $\chi_j$ can be determined in terms of $r_0$ and $\chi_s$, for any value of $\chi_s$. This leading order will necessarily come from the dominant terms (1st, 2nd and 4th) of (\ref{modTaubeschi}). Therefore, we look for the order of the resonance as follows. Keep just these dominant terms and substitute $\chi=\left(\frac{\Omega(r_0)}{p}\right)^{p/2}(r-r_0)^p+\chi_{s}(r-r_0)^{p+s}$. Expand the resulting expression in powers of $r-r_0$, keeping only the leading order of terms involving $\chi_s$, which will clearly be linear in $\chi_s$ (because $s>0$):
$$
(r-r_0)^{p+s-2}(s^2-s-2)\chi_s,
$$
whose vanishing implies $s=-1$ or $s=2$. The second and positive root indicates a resonance at order $p+2$ of the expansion in $r-r_0$, which means that $\chi_2$ can only appear in the coefficient of order $(r-r_0)^{p+1}$ or higher in the expansion of (\ref{modTaubeschi}) when $\chi$ is replaced by the power series. But at these orders the coefficients $\chi_{j\geq 3}$ are present and thus $\chi_2$ is not fixed. 

We are interested in analysing the order of resonance, as this will provide constraints on the geometry of $\Sigma$. Thus, we write $\chi=\left(\frac{\Omega(r_0)}{p}\right)^{p/2}(r-r_0)^p+\chi_1(r-r_0)^{p+1}+\chi_2(r-r_0)^{p+2}$, substitute it in (\ref{modTaubeschi}) and divide by $(r-r_0)^p$ to get
\begin{align}\label{PAeq}
&(r-r_0)^p \left(\frac{-2 (p-1) p^{-\frac{p}{2}}  \Omega (r_0)^{p/2}\chi_1-\Omega (r_0)^{p-1} \Omega '(r_0)}{r-r_0}-\frac{p^{1-p} \Omega (r_0)^p}{(r-r_0)^2}\right)+\nonumber\\
&+\frac{1}{r-r_0}\left(p^{1-\frac{p}{2}} \Omega (r_0)^{\frac{p}{2}-1} \Omega '(r_0)+\frac{p^{1-\frac{p}{2}} \Omega (r_0)^{p/2}}{r_0}-2 \chi_1\right)+ (p-2)  \Omega(r_0)^{-1} \Omega '(r_0)\chi_1+\nonumber \\ 
&+\frac{p+1}{r_0}  \chi_1-2 (p-1) p^{p/2-1} \Omega (r_0)^{-p/2}\chi_1^2+\frac{p^{-p/2+1}}{2} \Omega(r_0)^{p/2-1} \Omega''(r_0)-\frac{p^{-p/2+1}}{r_0^2}  \Omega(r_0)^{p/2},
\end{align}
up to order of a positive power of $r-r_0$.

Equating the coefficient of each order to zero, we find conditions on $\Omega''(r_0)$ and $\chi_1$. As mentioned above, had we written $\chi$ as an infinite series $\chi=\sum_{n\geq 0}\chi_n (r-r_0)^{p+n}$, we would have been able to calculate recursively $\chi_n\;(n\geq 3)$ in terms of $r_0$ and $\chi_2$. 
 
If $p\geq 2$, we calculate $\chi_1$ from the term of order $(r-r_0)^{-1}$ and then the term of order $(r-r_0)^0$ gives the following equation for the conformal factor
$$
\Delta_0 \ln \Omega(r_0)=0,
$$
which is valid for any $r_0$, yielding a differential equation whose solution is $\Omega(r)=c_1 r^{c_2}$, for some constants $c_1>0$ and $c_2>-2$, thus the surface $\Sigma$ is locally flat. We require that $c_2>-2$ since the origin $r=0$ would be at infinite distance from any other point otherwise. In fact, by performing the change of radial variable $R=\frac{2\sqrt c_1}{c_2+2}r^{\frac{c_2+2}{2}}$, the metric becomes $dR^2+\left(\frac{c_2+2}{2}\right)^2R^2d\theta^2$ so that we can set $\Omega=1$ in (\ref{modTaubeschi}) at the expense of introducing a deficit at the angular variable $\theta$, characterising a conical singularity at the origin. We suppose that the background is a smooth manifold and therefore we do not take into account these singularities and suppose $c_2=0$.

If $p=2$, the term of order $(r-r_0)^{p-2}$ in (\ref{PAeq}) contributes. Thus, the vanishing of (\ref{PAeq}) at orders $(r-r_0)^{-1}$ and $(r-r_0)^0$ implies $\chi_1= \frac{1}{2}\left(\Omega'(r_0)+\frac{\Omega(r_0)}{r_0}\right)$ and 
\begin{equation}
\Delta_0 \ln \Omega(r_0)=\Omega(r_0).
\end{equation}
This equation means that $\Sigma$ has constant Gauss curvature $-1/2$. This is not surprising as for $p=2$, $q=1$ thus equation (\ref{modTaubes}) is the usual Taubes equation, up to replacing $h$ by $-h$, whose Painlev\'e integrability was studied in \cite{Shiff1991}. Solutions to the modified Taubes equation in this case would involve a Blaschke product but from condition (\ref{eqBCond}) the magnetic field $B$ would not be integrable due to divergences where the Higgs field vanishes and thus we would not be able to define a magnetic flux and the energy would be infinite. Let us however point out that this case admits the following solutions 
\begin{align}
\chi&=\frac{4 r^2 (\ln r)^2}{(1-r^2)^2}, \label{solLiou1}\\
\chi&=\frac{(r^{c+1}-r^{-c+1})^2}{c^2(1-r^2)^2}, \quad 0<c<1,\label{solLiou2}\\
\chi&=\frac{4 r^2 \sin^2(c \ln r)}{c^2(1-r^2)^2}, \quad c>0,\label{solLiou3}
\end{align}
which are not analogous to Abelian vortices on smooth surfaces and can be ruled out by the conditions imposed in the end of section \ref{secModifiedVortices}. These solutions were obtained from results of \cite{Popov1993} (cf. also section $5$ of \cite{ConDor2015}).

If $p=1$ then all the terms in (\ref{PAeq}) contribute and we find that the conformal factor should satisfy the following differential equation
\begin{equation}\label{conffactorpeq1}
\Omega''(r_0)-\frac{\Omega'(r_0)^2}{\Omega (r_0)}+\frac{\Omega '(r_0)}{r_0}-\Omega '(r_0) \sqrt{\Omega (r_0)}-\frac{2 \Omega (r_0)^{3/2}}{r_0}=0,
\end{equation}
which can be rewritten in terms of $F(r)=\sqrt{\Omega(r)}$ as
\begin{equation}\label{diffeq}
F''(r)-\frac{F'(r)^2}{F(r)}+\frac{F'(r)}{r}-\frac{F(r)^2}{r}-F(r) F'(r)=0.
\end{equation}
Its general solution is 
\begin{equation}\label{eq1}
F(r)=\frac{C_1}{r} \frac{(C_2 r)^{C_1}}{1-(C_2 r)^{C_1}},
\end{equation}
where $C_1$ and $C_2$ are arbitrary positive constants, so that the origin $r=0$ is at finite distance from any other point. Under the change of variables $R=(C_2 r)^{C_1/2}$, equation (\ref{modTaubeschi}) becomes
\begin{equation}\label{modTaupeq1}
\dfrac{d^2\chi}{dR^2}-\frac{1}{\chi}\left(\dfrac{d\chi}{dR}\right)+\frac{1}{R}\dfrac{d\chi}{dR}+\frac{4 R^2}{(1-R^2)^2}\frac{1}{\chi}\left(1-\chi\right)=0,
\end{equation}
where we assume that $0\leq R<1$. A one parameter family of solutions to this equation was given in \cite{Shiff1991} (cf. equations (2.15--2.16) of this reference).
It satisfies the necessary conditions for Painlev\'e property established so far, but its analysis is not finished yet. To complete Painlev\'e test we follow the usual procedure. Expand $\chi=\chi_0 (R-R_0)+\chi_1 (R-R_0)^2+\chi_2 (R-R_0)^3+\cdots$, substitute it in (\ref{modTaupeq1}) and expand the left hand side in powers of $R-R_0$. The vanishing of the leading order implies
$$
\chi_0=\pm\frac{2 R_0}{1-R_0^2},
$$
where $0\leq R_0 <1$.
The case in which we choose the $+$ sign was already analysed above and led us to equation (\ref{conffactorpeq1}). Now, if we choose the $-$ sign, the vanishing of the new leading term implies 
$$
\chi_1=-3\frac{1+R_0^2}{(1-R_0^2)^2}.
$$
With these choices of $\chi_0$ and $\chi_1$, the left hand side of (\ref{modTaupeq1}) becomes
$$
-\frac{16 R_0}{(1-R_0^2)^3}(R-R_0)+O\left((R-R_0)^2\right),
$$
whose first term cannot be eliminated by any choice of $\chi_i$. This means that the expansion of $\chi$ should involve logarithmic terms of the form $\ln(R-R_0)$. Therefore, equation (\ref{modTaupeq1}) does not pass the Painlev\'e test. Another solution to (\ref{diffeq}) can be obtained by taking the limit $C_1\to 0$ in (\ref{eq1}) and then equation (\ref{modTaubeschi}) becomes
$$
\dfrac{d^2\chi}{dR^2}-\frac{1}{\chi}\left(\dfrac{d\chi}{dR}\right)+\frac{1}{R}\dfrac{d\chi}{dR}+\frac{ R^2}{(\ln R)^2}\frac{1}{\chi}\left(1-\chi\right)=0,\quad R=C_2 r,
$$
which also fails the Painlev\'e test as a similar calculation shows.

In the case $p<0$, a similar procedure will lead to the condition $p=-\frac{4}{1-q}$. The conditions for Painlev\'e integrability can be derived from the case $p>0$ above. In fact, under the change of variables $\chi\mapsto 1/\chi$ in equation (\ref{modTaubeschi}), $q$ is changed into $-q$ or, writing this equation in terms of $p$ using $q=\frac{4-|p|}{p}$, $p$ is changed into $-p$. Therefore, the conditions for Painlev\'e integrability in the cases $p=-1$, $p=-2$ and $p\leq -3$ are the same as in the cases $p=1$, $p=2$ and $p\geq 3$, respectively. Namely, for $p=-1$, there are no integrable soliton solutions, for $p=-2$, $\Sigma$ must be a hyperbolic space with constant curvature $-1/2$ and for $p<-2$, $\Sigma$ must be flat up to conical singularities. Even though from the integrability point of view cases $p=-2$ (or $q=-1$) and $p=2$ (or $q=1$) are the same, for $p=-2$ we obtain the ordinary Taubes equations of the Abelian Higgs model, which admits soliton solutions satisfying our conditions as opposed to the case $p=2$.

To complete the integrability analysis we need to have a closer look in the range $|p|\geq 3$, in which case $-1/3 \leq q \leq 1/3$. This is because for this range of $q$, we can find a $p_1>0$ and a $p_2<0$ such that $q=\frac{1}{p_i}\left(4-|p_i|\right),\; i=1,2$. But for Painlev\'e integrability to take place, the integrability conditions should hold for all possible choices of leading order $p$. Therefore, we have to solve 
$$
\frac{1}{p_1}\left(4-p_1\right)=q=\frac{1}{p_2}\left(4+p_2\right),
$$
for integers $p_1>0$ and $p_2<0$. There are exactly three solutions to this equation: $(p_1,p_2)=(6,-3),\;(4,-4)$ and $(3,-6)$, yielding $q=-\frac{1}{3},\;0$ and $\frac{1}{3}$, respectively. Since the cases $q=1/3$ and $q=0$ lead to the models studied in \cite{MD2012}, we will present the explicit vortex solutions to the case $q=-1/3$ in section \ref{solutions} (cf. equation (\ref{solMik})), after the Painlev\'e analysis of the PDE.

\section{Painlevé analysis of the PDE}\label{secPDE}

We will find all possible choices of $G(e^{h/2})^2=e^{(q+1)h/2}$ and of background metric $\Omega$ such that equation (\ref{modTaubes}) admits has the Painlev\'e property, now without imposing any symmetry to the PDE. We will do the analysis using the method proposed by Weiss, Tabor and Carnevale \cite{WTC1983}, which is the analogue for PDEs. As in the previous section, in order to avoid the logarithmic singularities in the analysis we look instead at the equation for $\chi=e^{h}$,
\begin{equation}\label{PDEChi}
\Delta_0\chi-\frac{1}{\chi}|\nabla\chi|^2+\frac{\Omega(x,y)}{\chi^{(q-1)/2}}\left(1-\chi\right)=0,
\end{equation}
where $\nabla\chi=(\partial_x\chi,\partial_y\chi)$ is the gradient vector of $\chi$ and $|\nabla\chi|^2=\left(\partial_x\chi\right)^2+\left(\partial_y\chi\right)^2$ is its Euclidean norm.

We look for the dominant behaviour by setting 
$$
\chi\sim\chi_0(x,y)\varphi(x,y)^p,
$$
where $\chi_0$ is a non-zero function to be determined and $p$ is an integer, as justified bellow. Keeping the lowest order terms in $\varphi$, we find
\begin{equation}\label{eqloworders}
\chi_0^{(1-q)/2}\Omega\,(1-\chi_0\,\varphi^p)\, \varphi^{p(1-q)/2}-p\, \chi_0 \,|\nabla\varphi|^2\, \varphi^{p-2}=0.
\end{equation}
We then need to separate the analysis into two different cases, $p>0$ and $p<0$. 

If $p>0$ then the term in $\varphi^p$ in the parenthesis of (\ref{eqloworders}) is of higher order and can be neglected at this stage. Then we equate the powers of $\varphi$ for the remaining two terms, $p(1-q)/2=p-2$ which gives a relation between $q$ and $p$, which  will be convenient to be solved for $q$:
$$
q=\frac{4-p}{p}.
$$
We solve (\ref{eqloworders}) for $\chi_0$ to find 
$$
\chi_0=\left(\frac{\Omega}{p |\nabla\varphi|^2}\right)^{p/2}.
$$

Anticipating from the ODE analysis above that there will be a resonance at second order, we expand $\chi$ as $\chi=\varphi^p(\chi_0+\chi_1 \varphi+\chi_2 \varphi^2)$, substitute it in (\ref{PDEChi}), divide by $\varphi^p$ and expand the whole expression in powers of $\varphi$ up to the first two lowest orders, which are $\varphi^{-1}$ and $\varphi^0$, keeping the terms of order $\varphi^{p-2}$ and $\varphi^{p-1}$. The terms of order $\varphi^{p-2}$ and $\varphi^{p-1}$ are 
\begin{equation}\label{termorderp}
-p^{1-p}\Omega^p |\nabla\varphi|^{2-2p}\varphi^{p-2}
\end{equation}
and
\begin{equation}\label{termorderp1}
-2(p-1)\left(\frac{\Omega}{p}\right)^{p/2} |\nabla\varphi|^{2-p}\chi_1\varphi^{p-1},
\end{equation}
respectively, which arise from the very last term in (\ref{PDEChi}).

These terms will not contribute to the analysis if $p>2$. Therefore, we will separate the analysis into the cases $p=1$, $p=2$ and $p\geq 3$.

If $p=1$, then the vanishing of the term of order $\varphi^{-1}$, which involves (\ref{termorderp}), gives rise to an algebraic equation for $\chi_1$ whose solution is
$$
\chi_1=\frac{\sqrt{\Omega(x,y)}}{2|\nabla\varphi|^3}\left(\Delta_0\varphi-\sqrt{\Omega(x,y)}|\nabla\varphi|\right).
$$
Then the term of order $\varphi^0$ will not depend on $\chi_2$, manifesting the resonance at this order predicted above. Instead, this term is a fairly big expression involving $\varphi$ and $\Omega$ (and their partial derivatives up to second order) that should vanish for any small function $\varphi$. Making the choices $\varphi=\pm\epsilon x$, $\varphi=\pm\epsilon y$ and $\varphi=\epsilon x y$, where $\epsilon$ is a small positive constant, we get differential equations for $\Omega$ that can only be solved by $\Omega=0$. This case is thus not interesting for our purposes.

If $p=2$ then $q=1$ and a change of variables of the form $\chi\mapsto\chi^{-1}$ will put (\ref{PDEChi}) in the form of the usual Taubes equation, whose Painlev\'e analysis requires $\Sigma$ to be a hyperbolic space of curvature $-1/2$ \cite{Shiff1991}. As for the ODE in the previous section, condition (\ref{eqBCond}) implies that the divergence of the magnetic field (\ref{modBog2}) at each zero of the Higgs field would make the magnetic flux infinite, and thus no solution would fit our requirements.

If $p\geq 3$ then the lowest order term is 
$$
\left(\frac{\Omega}{p|\nabla\varphi|^2}\right)^{p/2}\left[p\Delta_0\varphi-2\left(\frac{p}{\Omega}\right)^{p/2}|\nabla\varphi|^{p+2}\chi_1\right]\frac{1}{\varphi},
$$
The term in brackets should vanish, resulting in an equation for $\chi_1$ which can be solved by
$$
\chi_1=\frac{p^2}{2\Omega}\left(\frac{\Omega}{p|\nabla\varphi|^2}\right)^{(p+2)/2}\Delta_0\varphi.
$$
This choice of $\chi_1$ annihilates the term of order $\varphi^{-1}$ and we are left with the term of order $\varphi^0$  which is
$$
-\frac{p}{2}\Delta_0\ln\Omega \left(\frac{\Omega}{p|\nabla\varphi|^2}\right)^{p/2}.
$$
We notice that it does not involve $\chi_2$, indicating the resonance anticipated earlier. The conformal factor $\Omega$ should then satisfy $\Delta_0\ln\Omega=0$. In other terms, the metric should be flat, up to possible conical singularities. Thus, we can choose local coordinates to set $\Omega=1$ under smoothness assumptions. 

As for the ODE, the conditions for Painlevé integrability in the cases $p=-1$, $p=-2$ and $p\leq -3$ are the same as for $p=1$, $p=2$ and $p\geq 3$, respectively, as we can go from $p$ to $-p$ by changing $\chi$ into $\chi^{-1}$. Therefore, the integrable cases for the PDE correspond to the same as for the ODE, that is to say either $\Sigma$ is a hyperbolic surface of curvature $-1/2$ and $|q|=1$ or $\Sigma$ is flat and $|q|=1/3$. Notice however that for $p=1$ (or $q=1$)  we did not have a soliton solution but for $p=-1$ (or $q=-1$) we find exactly the usual Abelian Higgs model on hyperbolic surfaces, whose solutions are well understood.

Here it is worth pausing to explain why we require $p$ to be an integer. If $p$ is not an integer then the PDE does not admit the Painlev\'e property, however it may be transformed into one having this property under a change of variables replacing $\chi$ by some power of $\chi$, which might reveal further integrability properties. However, once we substitute the series expansion $\chi=\varphi^p\sum_{k\geq 0}\chi_k\varphi^k$ in (\ref{PDEChi}) and divide the left hand side by $\varphi^p$, the resulting expression takes the form
$$
\left(\text{power series in }\varphi\right) -\Omega \varphi^{p-2}\left(\sum_{k\geq 0}\chi_k\varphi^k\right)^{2\frac{p-1}{p}}=0,
$$
and for the second term to vanish for $p$ non-integer while $\chi_0\not=0$ we would need to require that $\Omega=0$, which is not of our interest.

We have done the Painlev\'e analysis by expanding the $\chi$ in power series of $\varphi$. We could have also used the ``reduced ansatz" proposed by M. Kruskal and explained in \cite{WTC1983} which consists in supposing that $\partial_x\varphi\not=0$ and expanding $\chi$ in power series of $x-\psi(y)$, where $\psi$ is a function such that $\varphi(\psi(y),y)=0$ that exists by the implicit function theorem. Even though this ansatz is clearly analogous to the Painlev\'e analysis for ODEs and can simplify calculations considerably, in our case we would have needed to expand $\Omega(x,y)$ in power series of $x-\psi(y)$ with respect to the first variable and thus we decided not to use it. Anyway, similar calculations with this ansatz yield the same results.

\subsection{Explicit solutions}\label{solutions}

For $\Omega=1$ and $q=0$, (\ref{PDEChi}) becomes the sinh-Gordon equation $\Delta_0 \frac{h}{2}=\sinh\frac{h}{2}$ while for $q=\pm \frac{1}{3}$, it becomes the Tzitz\'eica equation \cite{Mik1979,Mik1981,FordyGib1980}
\begin{equation}\label{Tzieq}
\Delta_0 u+\frac{1}{3}\left(e^{-2u}-e^u\right)=0,
\end{equation}
where $u=q h$. These equations were studied in the context of Abelian vortices in \cite{MD2012}, where the cases considered correspond to $q=\frac{1}{3}$ and $q=0$ in our language. However, the analysis presented here points to a new solution in the case $q=-\frac{1}{3}$ and completes the list of integrable cases under the class of models considered. We will focus on the details of this new solution, bearing in mind that it is analogous for the other two cases.

We still need to apply the boundary conditions so that we can calculate physical quantities such as the energy, magnetic flux and vortex strength. We thus have to know the behaviour of the asymptotics of the solutions to (\ref{Tzieq}). If we apply the cylindrical symmetry reduction $u=u(r)$, supposing that $u$ is only a function of the radial coordinate, (\ref{Tzieq}) reduces to a Painlev\'e III equation with choice of parameters $(1,0,0,-1)$ under the change of variables $u(r)=\ln w(r)-\frac{1}{2}\ln r+\frac{1}{4}\ln \frac{27}{4},\; r=\frac{3\sqrt{3}}{2}\rho^{2/3}$ :
$$
\dfrac{d^2 w}{d\rho^2}=\frac{1}{w}\left(\dfrac{dw}{d\rho}\right)^2-\frac{1}{\rho}\dfrac{dw}{d\rho}+\frac{w^2}{\rho}-\frac{1}{w}.
$$
The behaviour of its solutions in the asymptotics were studied in \cite{Kit1989}. We thus apply this reduction and equation (18) in \cite{Kit1989} with $g_1=g_2=0$, $g_3=1$, $\tau=\frac{r^2}{12}$ and $s=1+2\cos\left[\frac{\pi}{9}(6-2N)\right]$ to find
\begin{equation}\label{solMik}
h=-3u \sim_{r\to 0} -3\ln\left[\frac{2\alpha}{9}(N-3)^2 12^{\frac{N}{3}}\frac{r^{-\frac{2N}{3}}}{\left(1-\alpha 12^{\frac{1}{3}(N-3)}r^{-\frac{2}{3}(N-3)}\right)^2}\right],
\end{equation}
where
$$
\alpha=3^{\frac{2}{3}(N-3)}\frac{\Gamma\left(\frac{1}{3}\left(2+\frac{N}{3}\right)\right)\Gamma\left(\frac{1}{3}\left(1+\frac{2N}{3}\right)\right)}{\Gamma\left(\frac{1}{3}\left(4-\frac{N}{3}\right)\right)\Gamma\left(\frac{1}{3}\left(5-\frac{2N}{3}\right)\right)}
$$
and $N$ is the topological charge (or vortex number), which is allowed to take values $N=1$ and $N=2$.

The results below fig. 1 in the same reference gives the behaviour at $r\to\infty$
\begin{equation}\label{solMikinfty}
h=-3u \sim_{r\to\infty}-\frac{3\sqrt{3}}{\pi}\left\{1+2 \cos\left[\frac{\pi}{9}\left(6-2N\right)\right] \right\}K_0(r),
\end{equation}
were $K_0(r)\sim_{r\to\infty}\sqrt{\frac{\pi}{2r}}e^{-r}$ is the modified Bessel function of second kind. The strength of the vortex can be read off from the coefficient before the Bessel function $K_0$ and takes approximate values $2.23$ and $4.19$ for $N=1$ and $2$, respectively. For comparison, these values are approximately $1.80$ and $1.45$ for the models with $q=0$ and $q=1/3$, respectively, where only $N=1$ vortex solutions are allowed \cite{MD2012}.

In Figure \ref{plots} we plot the magnitude of the Higgs field square $|\phi|^2$ and the magnetic field $B$ as functions of $r$ associated to this solution for both vortex numbers. We notice, using equation (\ref{solMik}), that the magnetic field blows up at the origin as $B\sim r^{-2N/3}$ and would not be integrable for $N\geq 3$. This restricts $N$ to be $1$ or $2$, as mentioned above, in order to obtain a finite magnetic flux. In fact, a direct calculation shows that $\int_{\Sigma} B =2\pi N$ (cf. also equation (\ref{modEn})). It can be done by using equations (\ref{modBog2}) and (\ref{modTaubes}) along with rotational symmetry to write $B=-\frac{1}{2r}\dfrac{d}{dr}\left(r\dfrac{dh}{dr}\right)$, then
$$
\frac{1}{2\pi}\int_{\Sigma}B=\frac{1}{2\pi}\int_0^{2\pi}\int_0^\infty-\frac{1}{2r}\dfrac{d}{dr}\left(r\dfrac{dh}{dr}\right) r dr d\theta=-\frac{1}{2}\left[r\dfrac{dh}{dr}\right]_{r=0}^{\infty}=N,
$$
where we have performed an integration by parts in the second equality and used the asymptotic expressions (\ref{solMik}) and (\ref{solMikinfty}) in the last one.

The magnetic field for the models corresponding to $q=0$ and $q=1/3$ present a similar behaviour. At the origin they diverge as $B\sim r^{-1}$ and $B\sim r^{-4/3}$, respectively, while they monotonically tend to zero at infinity. Both give the same magnetic flux $2\pi$ corresponding to $N=1$ vortex solutions.

\begin{center}
\vspace{-3.5cm}
  \begin{figure}[ht]
  \begin{tabular}{ll}
\hspace{-2.2cm}\includegraphics[scale=0.5]{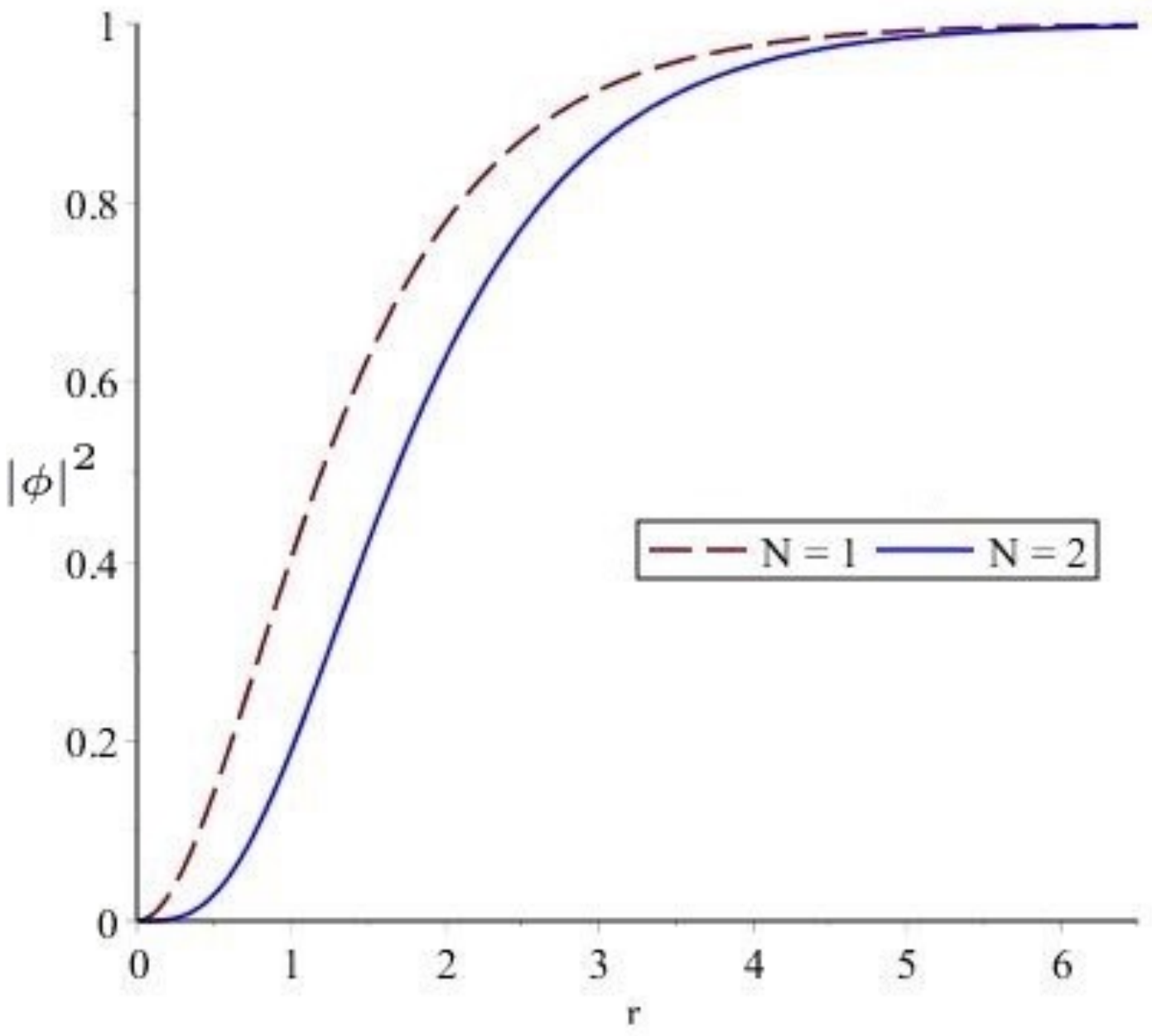} &\hspace{-2cm} \includegraphics[scale=0.5]{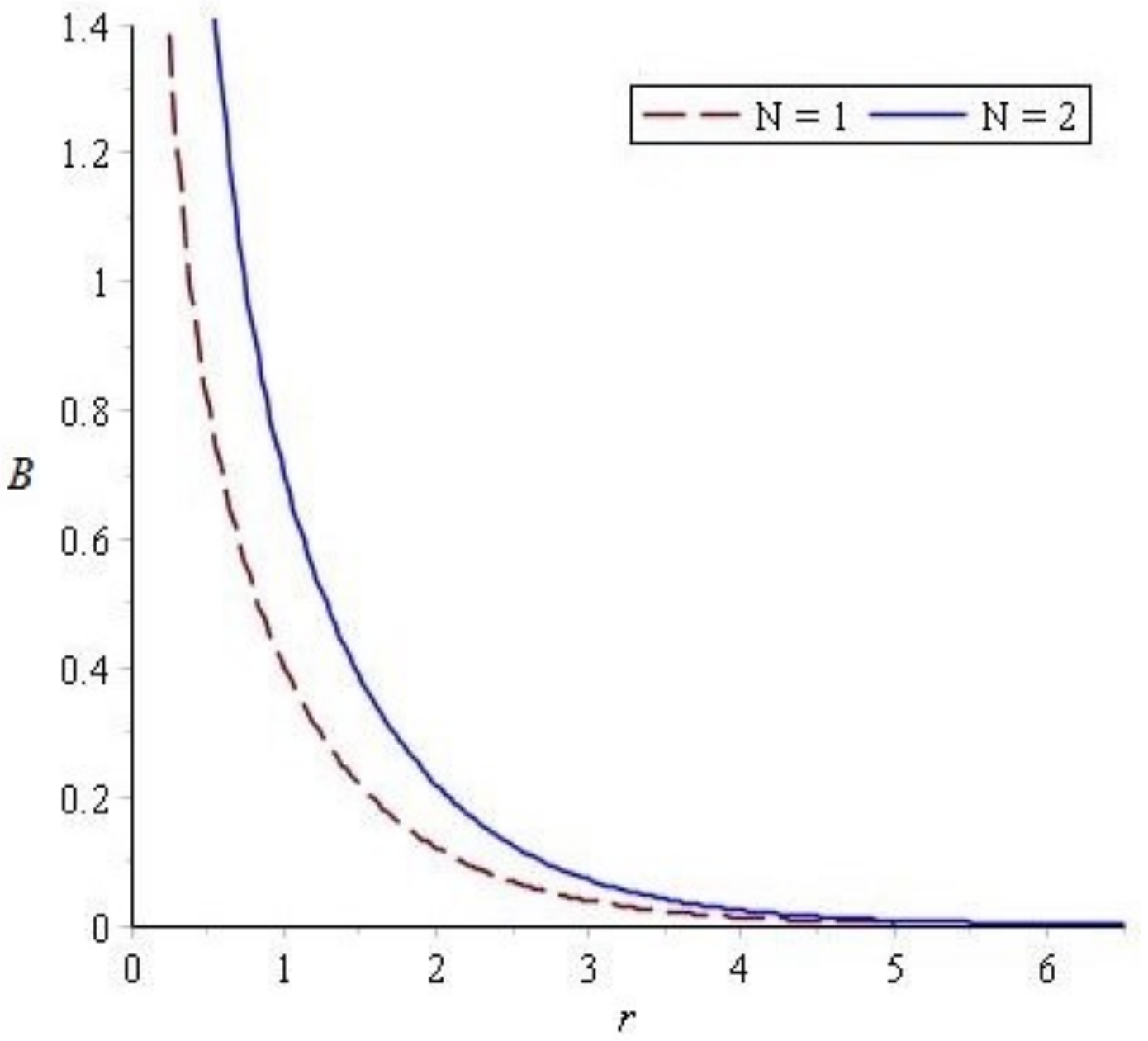}
  \end{tabular}
  \vspace{-4cm}
\caption{Square of the modulus of the Higgs field $|\phi|^2$ (left plot) and the magnetic field $B$ (right plot) as functions of $r$ for vortex number $N=1$ (dashed lines) and $N=2$ (full lines) for solution (\ref{solMik} --\ref{solMikinfty}).}
  \label{plots}
  \end{figure}
\end{center}

\section{Conclusion}

The solutions presented here relied on the ansatz $G(e^{h/2})^2=e^{(q+1)h/2}$, which was inspired by particular cases of integrable Abelian-Higgs vortices on non-hyperbolic backgrounds \cite{MD2012}. Also, equation (\ref{modTaubes}) suggests that these solitons can be interpreted as usual Abelian-Higgs vortices on the background $\Omega/G^2$ \cite{MD2012}. Under this framework, the ansatz for $G$ is a natural generalisation of the metric one obtains by multiplying the conformal factor of the background metric by an integer power of the absolute value of the Higgs field in the non-linear superposition rule of vortices \cite{Baptista2014}. Nonetheless, it may be worth noticing that other choices of $G$ that were not explored here would also lead to integrable vortices. For instance, choosing $G(e^h)^2=-\frac{1-e^h}{h}$ and $\Omega=1$ and imposing cylindrical symmetry would give rise to a modified Bessel equation, or we could find solutions in terms of hypergeometric functions by choosing $G^2=1-e^h$ and $\Omega=e^{-r}$. Clearly, these two cases would not arise from the Bogomolny argument we used as it requires $G^2>0$ for any $h$. 

It is interesting to notice that modifying the model may bring to light further integrable cases that could not be found otherwise by standard methods of integrability, even though these other cases might be isolated and not involving any moduli. Moreover, using results of \cite{Popov2009} these vortices give rise to cylindrically symmetric instantons on a $4$-dimensional background that is not (anti-)self-dual \cite{ConDor2015}.

\section*{Acknowledgements}
I am grateful to Maciej Dunajski, Nick Manton, Daniele Dorigoni, Raphael Maldonado and Mark Ablowitz for helpful discussions and to Cambridge Commonwealth, European \& International Trust and CAPES Foundation Grant Proc. BEX 13656/13-9 for financial support.
\bibliography{BiblioArticle} 
\bibliographystyle{plain}

\end{document}